\documentclass[12pt,english]{revtex4-1}
\usepackage[T1]{fontenc}
\usepackage[latin9]{inputenc}
\usepackage{amsmath}
\usepackage{amssymb}
\usepackage{graphicx}
\usepackage{setspace}
\usepackage[hidelinks]{hyperref}

\makeatletter

\usepackage{lettrine}
\usepackage{babel}

\makeatother

\usepackage{babel}
\begin{document}

\title{Quantum Annealing for Combinatorial Clustering}

\author{Vaibhaw Kumar}
\email{kumar_vaibhaw@bah.com}

\selectlanguage{english}%

\author{Gideon Bass}
\email{bass_gideon@bah.com}

\selectlanguage{english}%

\author{Casey Tomlin}
\email{tomlin_casey@bah.com}

\selectlanguage{english}%

\author{Joseph Dulny III}
\email{dulny_joseph@bah.com}

\selectlanguage{english}%

\affiliation{Booz Allen Hamilton ~\\
901 15$^{\mathrm{th}}$ Street NW, Washington, D.C. 20005}

\date{\today}
\begin{abstract}
\singlespacing
Clustering is a powerful machine learning technique
that groups ``similar'' data points based on their characteristics.
Many clustering algorithms work by approximating the minimization
of an objective function, namely the sum of within-the-cluster
distances between points. The straightforward approach involves examining
all the possible assignments of points to each of the clusters. This
approach guarantees the solution will be a global minimum, however
the number of possible assignments scales quickly with the number
of data points and becomes computationally intractable even for very
small datasets. In order to circumvent this issue, cost function minima
are found using popular local-search based heuristic approaches such
as $k$-means and hierarchical clustering. Due to their greedy nature,
such techniques do not guarantee that a global minimum will be found
and can lead to sub-optimal clustering assignments. Other classes
of global-search based techniques, such as simulated annealing, tabu
search, and genetic algorithms may offer better quality results but
can be too time consuming to implement. In this work, we describe
how quantum annealing can be used to carry out clustering. We map
the clustering objective to a quadratic binary optimization (QUBO)
problem and discuss two clustering algorithms which are then implemented
on commercially-available quantum annealing hardware, as well as on
a purely classical solver ``qbsolv.'' The first algorithm assigns
$N$ data points to $K$ clusters, and the second one can be used
to perform binary clustering in a hierarchical manner. We present
our results in the form of benchmarks against well-known $k$-means
clustering and discuss the advantages and disadvantages of the proposed
techniques.
\end{abstract}
\maketitle

\section{\label{sec:intro}Introduction}

\label{intro}
Clustering is the process of grouping objects based on their common
features. It is a powerful machine learning technique used to digest
and interpret data. Clustering algorithms find application in a wide
variety of fields, including the investigation of gene expression
patterns \cite{BenDor1999,Das2016,gorzalczany2016,marisa2013}, document
clustering \cite{Xie2013,Balabantaray2015}, and consumer segmentation
\cite{mudambi2002,sharma2013,chan2012}, among many others.

Often, clustering is cast as an optimization problem using an objective
(or `cost') function \cite{friedman2001elements}. A common objective function is the sum over
pairwise dissimilarities within the clusters:

\begin{equation}
W(C)=\frac{1}{2}\sum_{a=1}^{K}\sum_{C(i)=C_{a}}\sum_{C(i')=C_{a}}d(x_{i},x_{i'})\label{eq:objective-1}
\end{equation}
Here each $x_{i}$ represents an individual data point or observation,
$d$ is a distance metric on the space of possible data points, and
$C$ refers to the cluster assignment. Essentially, this results in
a value given by the sum of the total distances between all points
that reside in the same cluster. In combinatorial clustering, the
total dissimilarity of every possible cluster assignment is examined
to find the minimum of $W$, which guarantees a global minimum is
found. Unfortunately, examining all possible assignments of observations
to clusters is feasible only for very small datasets. For a task where
$N$ data points are assigned to $K$ clusters, the number of distinct
assignments is given by

\begin{equation}
S(N,K)=\frac{1}{K!}\sum_{i=1}^{K}(-1)^{K-i}{N \choose i}i^{N},\label{eq:num_assignments}
\end{equation}
and for example, $S(19,4)\approx10^{10}$ \cite{friedman2001elements}. Even with modern computing
resources, this exhaustive search technique quickly becomes impractical
or outright impossible.

Common clustering algorithms, such as $k$-means \cite{hartigan1979}
or hierarchical clustering \cite{johnson1967} are local search techniques
which take less exhaustive, greedy approaches. For a nice summary
of how clustering techniques have evolved, see \cite{jain2010}.
Because they are heuristic-based greedy approaches, neither $k$-means
nor hierarchical clustering are guaranteed to find an optimal clustering.
A single run of the $k$-means algorithm is likely to end up with
a locally optimal solution that is not a global optimum. Thus, $k$-means
is typically run multiple times with different random initializations
to get a subset of good solutions. The best solution is then chosen
as the final clustering result. However, even with many runs, there
is no guarantee that the true global minimum will be found, particularly
if the solution space has many local minima.

The problem of choosing an optimal solution from many potential combinations,
or combinatorial optimization, is well known in the computer science
literature. Many formulations of the problem, including clustering,
are NP-hard \cite{Garey1979}. These problems, including the famous
traveling salesman problem, suffer from similar scaling issues \cite{papadimitriou1977}.
For clustering related problems, global search techniques such as
simulated annealing, tabu search, and genetic algorithms although
computationally less efficient than $k$-means are known to produce better quality
results \cite{al1996computational}.

We discuss the simulated annealing (SA) \cite{kirkpatrick1983} based
approach in a bit more detail, as it closely resembles the techniques
we describe in this paper, and it has been widely used for clustering
\cite{selim1991}. SA can be understood as analogous to the metallurgical
process from which its name derives\textemdash slowly cooling a metal.
During SA, the objective function is reimagined as the energy of a
system. At all times, there are random transitions that can potentially
occur. Which transitions actually occur is a function of both temperature
and whether the transition would result in an increase or decrease
in energy. At high temperatures, changes that increase or decrease
the energy are almost equally likely, but as the temperature slowly
decreases, energy increases become less and less likely, and eventually
the system settles into a very low energy state.

In the context of clustering, SA starts with a random assignment of
observations to different clusters. In each iteration, one observation
is randomly chosen to be reassigned to a new cluster. The change in
the objective function, or energy, $\Delta W=W_{\mathrm{new}}-W_{\mathrm{old}}$,
caused by the reassignment is computed. If the energy decreases ($\Delta W<0$),
the new assignment is accepted; if it increases ($\Delta W>0$), the
new assignment is accepted with a probability proportional to $\mathrm{exp}(-\Delta W/T)$
where $T$ represents the ``temperature'' of the system. The process
is continued for a predetermined number of steps, during which the
temperature is reduced in some prescribed manner \cite{mitra1985}.
At relatively high temperatures, the algorithm can easily increase
its energy, escaping local minima. However, at low temperatures it
steadily decreases in energy, eventually reaching a minimum. If the
temperature is brought down slowly enough, the probability of arriving
at the globally optimal clustering assignment approaches one.

SA-based clustering has shortcomings as well. It works well for moderate-sized
data sets where the algorithm can traverse the energy landscape quickly.
However, the process becomes extremely sluggish for large datasets,
especially at low temperatures, and it becomes more likely to get
stuck in local minima. In addition, there are several free parameters,
and selecting a suitable cooling schedule is difficult and requires
a time-consuming trial and error approach tuned to the problem at
hand. There have been many attempts at improving the speed and quality
of SA results \cite{szu1987,ingber1989,bouleimen2003}, generally
by doing many sequential runs with a very short annealing schedule.

\subsection{Quantum annealing}

\label{intro:1}
Quantum annealing (QA) is an outgrowth of simulated annealing that
attempts to use quantum effects to improve performance \cite{kadowaki1998}.
In QA, quantum fluctuations are used to change energy states, instead
of SA's thermal excitations. This has been thought to improve performance
\cite{santoro2006}, and there have been some experimental results
with recent commercially available devices to this effect \cite{denchev2016}.

In the adiabatic limit as the transformation from $H_{\mathrm{i}}$
to $H_{\mathrm{f}}$ slows sufficiently, the quantum annealer is guaranteed
by the adiabatic theorem to finish in the ground state (if it begins
in the ground state) \cite{Born1928,AlbashLidar2016}. In practice,
real machines operate at finite temperature and it is impractical
to set arbitrarily long annealing times. Moreover, the required annealing
time is proportional to the spectral energy gap between the ground
and first excited state, something that is rarely known a priori.
Therefore it is common practice to perform multiple runs on the annealing
device, each with a short (on the order of microseconds) annealing
time, after which the solution and resulting energy is saved. After
a large number of runs, the one with the lowest overall energy is
selected as an approximation of the global minimum.

In order to use QA for optimization, one has to find a suitable mapping
of the problem's objective function to the energy states of a quantum
system. Quantum annealing devices first initialize a quantum system
in an easy-to-prepare ground state of an initial Hamiltonian $H_{\mathrm{i}}$,
which is then slowly evolved into a final Hamiltonian $H_{\mathrm{f}}$
whose ground state corresponds to the solution of the optimization
problem at hand; for example, 
\begin{equation}
H(t)=(1-s(t))H_{\mathrm{i}}+s(t)H_{\mathrm{f}},\qquad0\le t\le T\label{eq:adham}
\end{equation}
where $s$ is monotonically increasing with $s(0)=0$ and $s(T)=1$.

It has recently become possible to encode and solve real-life optimization
problems on commercially available quantum annealing hardware \cite{biamonte2016quantum}.Recent studies have demonstrated how QA can be used to carry out important machine learning tasks \cite{dulny2016developing,neven2009nips,denchev2013binary,farinelli2016quantum} including clustering \cite{kurihara2009quantum,sato2013quantum}. However, authors in Ref \cite{kurihara2009quantum,sato2013quantum} discuss a different class of "soft "clustering algorithms where a data point is probabilistically assigned to more than one cluster. Note that the algorithms presented in this work belong to the class of "hard" clustering algorithms where a data point is assigned to one and only one cluster. 

Current generations of these devices are designed to solve problems cast into
the form of an Ising spin glass \cite{Ising1925}: 
\begin{equation}
H_{\text{f}}=\sum_{i}h_{i}\sigma_{i}+\sum_{ij}J_{ij}\sigma_{i}\sigma_{j}\label{eq:quboham}
\end{equation}
Here $\sigma_{i}$ represents the state of the $i^{t\text{h}}$ qubit
(or the $z$-component of the $i^{\text{th}}$ Pauli spin operator),
and can take values $\pm1$, while $h_{i}$ and $J_{ij}$ are the
control parameters of the physical system and represent the bias on
each of the qubits, and coupling between two qubits, respectively.
A very similar problem, quadratic unconstrained binary optimization
(QUBO), is also commonly used as a template. Ising problems can be
trivially converted into QUBO and vice versa. In the next section
we describe how the clustering objective function can be recast as
a QUBO.

\section{Quantum Annealing Methods}
\label{sec:Methods}
\subsection{One-hot encoding}
\label{sec:one-hot}
Suppose $N$ points $\{x_{i}\}_{i=1}^{N}$ are to be assigned to $K$
clusters $\{C_{a}\}_{a=1}^{K}$. Let each point $x_{i}$ be associated
with a Boolean variable $q_{a}^{i}$ which indicates whether the point
is in cluster $C_{a}$ or not. We refer to this as a ``one-hot'' encoding,
familiar from the QUBO solution of the map coloring problem \cite{Dahl2013}.
Given the $\tfrac{1}{2}N(N-1)$ real, positive-valued separations
$d(x_{i},x_{j})$, the clustering objective $W(C)$ can be written
as 
\begin{equation}
H'=\tfrac{1}{2}{\textstyle \sum_{i,j=1}^{N}}d(x_{i},x_{j}){\textstyle \sum_{a=1}^{K}}q_{a}^{i}q_{a}^{j},\label{eq:ham_unconstrained}
\end{equation}
assuming one can guarantee that for each $i,$ only a single $q_{a}^{i}=1$
for some cluster label $a$, the rest being zero. Energetically-speaking
however, with this form of $H'$ the guarantee will not be honored,
since the most beneficial values the $q$ may take are all zeroes
(or at least such that $H'$ remains zero). Thus we add constraints
(one for each value of $i$) to the objective function in the form:
\begin{equation}
\phi^{i}=\left({\textstyle \sum_{a=1}^{K}}q_{a}^{i}-1\right)^{2}\label{eq:constraint}
\end{equation}
times Lagrange multipliers $\lambda_{i}$ to $H'$: 
\begin{align}
H & :=H'+{\textstyle \sum_{i=1}^{N}}\lambda_{i}\phi^{i}\\
& ={\textstyle \frac{1}{2}}{\textstyle \sum_{i,j=1}^{N}}d(x_{i},x_{j}){\textstyle \sum_{a=1}^{K}}q_{a}^{i}q_{a}^{j}+{\textstyle \sum_{i=1}^{N}}\lambda_{i}\left({\textstyle \sum_{a=1}^{K}}q_{a}^{i}-1\right)^{2},
\label{eq:ham_constrained}
\end{align}
In practice, we must ensure that the $\lambda_i$ are chosen large enough
to discourage any constraint violation and force the one-hot encoding. In particular, we must protect against the mildest constraint violation, for which there are two possibilities: either a single $x_{i}$
can be assigned to more than one cluster, or it can be assigned to zero clusters.

If a point is assigned to more than one cluster, this can only result in an increase in $H$, simply because additional qubits being ``switched on'' can only introduce additional positive
terms in the sum involving same-cluster distances $d(x_{i},x_{j})$. Therefore, this type of violation will never be favored energetically. 

On the other hand, assigning $x_{i}$ to no cluster has the
effect of setting some $d(x_{i},x_{j})$ to zero, thus reducing $H$. The goal then is to choose the corresponding $\lambda_i$ large enough so that the constraint term offsets the maximum possible reduction in $H$. When a point $x_{i}$ is assigned no cluster, in the worst-case scenario it can set $N-K$ pairwise-distances to zero (because $x_{i}$ could cluster with a maximum of $N-K$ points, assuming every cluster contains at least one point; we take as given that $N\geq K+1$). The maximum possible reduction in $H$ would result when each of
the other points are at the maximum distance from $x_{i}$. In this
case $H$ is reduced by

\begin{equation}
\sum_{x_{j}\in C_{a}}d(x_{i},x_{j})\leq(N-K)\cdot\max_{x_{j}\in C_{a}}(d(x_{i},x_{j}))
\end{equation}

where the right-hand side is the worst-case scenario bound. Thus to guarantee that no constraint is violated, we set $\lambda_i = \lambda$ for all $i$ with

\begin{equation} 
\lambda_i \geq(N-K)\tilde{d},
\label{eq:lambda}
\end{equation}

where $\tilde{d}$ is the pair-pointwise maximum of $d(x_{i},x_{j})$
for all $i,j$. In practice, one can normalize $d(x_{i},x_{j})$ so that $\tilde{d}=1$ in (\ref{eq:lambda}). Then setting $\lambda = (N-K)$ in (\ref{eq:ham_constrained}) will ensure that 
no violations occur.  

In practice, $\lambda$ cannot be made arbitrarily large. Ultimately this will make demands on one's hardware precision, given the details of the clustering problem at hand. In the discussion that follows we comment on the precision limitations imposed by current-generation quantum annealing hardware specific to the one-hot encoding clustering approach. 

If the hardware supports $n$-bit precision for the couplings and biases,
and we assume only (dimensionless) integer values for those coefficients,
then the coupling and bias values one might reliably set and distinguish
are 
\begin{equation}
\{-(2^{n-1}-1),-(2^{n-1}-2),\dots,-1,0,1,\dots,2^{n-1}-2,2^{n-1}-1\}.
\end{equation}
To make a comparison with the hardware we will be using, we transform into spin variables.

\subsubsection{Spin variables}
\label{spin}
To determine the couplings and biases one should use on the hardware
annealer, or simulate this system, we must transform into spin variables
$s_{a}^{i}=2q_{a}^{i}-1$, so set $q_{a}^{i}=\frac{1}{2}\left(s_{a}^{i}+1\right)$
in $H$ to obtain (we drop constants along the way) 
\begin{eqnarray*}
	H & = & \tfrac{1}{2}{\textstyle \sum_{a=1}^{K}\sum}{}_{i,j=1}^{N}d(x_{i},x_{j})\tfrac{1}{2}(s_{ia}+1)\tfrac{1}{2}(s+1)\\
	& = & \tfrac{1}{8}{\textstyle \sum}_{a=1}^{K}{\textstyle \sum}_{i,j=1}^{N}d(x_{i},x_{j})s_{a}^{i}s_{a}^{j}+\tfrac{1}{4}\lambda{\textstyle \sum_{i=1}^{N}}{\textstyle \sum_{a\neq b}}s_{a}^{i}s_{b}^{i}\\
	&  & +\tfrac{1}{4}{\textstyle \sum}_{i,j=1}^{N}d(x_{i},x_{j}){\textstyle \sum}_{a=1}^{K}s_{a}^{i}+\tfrac{1}{2}\lambda(K-2){\textstyle \sum} _{i=1}^{N}{\textstyle \sum}_{a=1}^{K}s_{a}^{i},
\end{eqnarray*}
where we have used $\sum_{a\neq b}\left(s_{a}^{i}+s_{b}^{i}\right)=2(K-1)\sum_{a}s_{a}^{i}.$
The double sums are only over unique pairs, so we may write (introducing
an overall scaling) 
\begin{eqnarray*}
	H & = & \tfrac{1}{2}{\textstyle \sum} _{a=1}^{K}{\textstyle \sum} _{(i,j)}d(x_{i},x_{j})s_{a}^{i}s_{a}^{j}+\lambda{\textstyle \sum} _{i=1}^{N}{\textstyle \sum} _{(a,b)}s_{a}^{i}s_{b}^{i}\\
	&  & +{\textstyle \sum} _{(i,j)}d(x_{i},x_{j}){\textstyle \sum} _{a=1}^{K}s_{a}^{i}+2\lambda(K-2){\textstyle \sum} _{i=1}^{N}{\textstyle \sum} _{a=1}^{K}s_{a}^{i}.
\end{eqnarray*}
We want to maximize $\lambda$ given the hardware constraints. When
$K>2$ and the hardware is precise to $n$-bits, we thus set 
\[
2\lambda(K-2)=2^{n-1}-1
\]
or 
\begin{equation}
\lambda=\frac{2^{n-1}-1}{2(K-2)}.\label{eq:LambdaDef}
\end{equation}
and $H$ becomes 
\begin{eqnarray*}
	H & = & \tfrac{1}{8}{\textstyle \sum} _{a=1}^{K}{\textstyle \sum} _{i,j=1}^{N}d(x_{i},x_{j})s_{ia}s_{ja}+\frac{2^{n-1}-1}{K-2}{\textstyle \sum} _{i=1}^{N}{\textstyle \sum} _{b>a=1}^{K}s_{ia}s_{ib}\\
	&  & +\tfrac{1}{8}{\textstyle \sum} _{i,j=1}^{N}d(x_{i},x_{j}){\textstyle \sum} _{a=1}^{K}\left(s_{ia}+s_{ja}\right)+\left(2^{n-1}-1\right){\textstyle \sum} _{i=1}^{N}{\textstyle \sum} _{a=1}^{K}s_{ia}.
\end{eqnarray*}
The scaling requirement for $d$ reads 
\[
\left\vert d\right\vert \leq\frac{2\left(2^{n-1}-1\right)}{(N-K)\left(K-2\right)}.
\]
For $n=6$ (which is the precision on current commercially available
quantum annealers), this is 
\[
\left\vert d\right\vert \leq\frac{62}{(N-K)\left(K-2\right)}.
\]
For the minuscule test case $N=4$ and $K=3,$ one just needs to have
the maximum $d\leq62,$ but the couplings involving $d$ appear as
$\frac{1}{8}d,$ so these are restricted to be $<8,$ which doesn't
leave much room for accuracy. Clearly things only get worse as $N$
grows. For this reason, current QA hardware with six bits of precision
may only reliably accommodate small clustering instances. In the next
section we describe another QUBO formulation of the clustering objective
which enables us to carry out binary clustering and removes the precision
limitations posed by the one-hot encoding technique.

\subsection{Binary Clustering}
\label{binary}

In order to develop a better intuition we use spin variables here.
Suppose $N$ points are to be assigned to $K=2$ clusters. Let each
point $x_{i}$ be associated with an Ising variable $s_{i}=\pm1$
which indicates whether the point is in cluster 1 ($s_{i}=1$) or
not ($s_{i}=-1$). We refer to this as binary clustering. Without
the one-hot constraint, the objective is simply

\[
H=\frac{1}{2}\sum_{i,j=1}^{N}d(x_{i},x_{j})s_{i}s_{j}.
\]
When $d(x_{i},x_{j})$ is large, $s_{i}$ and $s_{j}$ tend to adopt
opposite spins (and thus are assigned to different clusters), whereas
if $d(x_{i},x_{j})$ is small, points tend to adopt the same spin
(and thus are assigned to the same cluster). This approach allows
assignment of points to only $K=2$ clusters as opposed the one-hot
encoding case. However, since there are no constraints, the precision
issues faced by one-hot encoding are avoided. Moreover, it requires
a significantly smaller number of qubits (half if we compare with
$K=2$ one-hot encoding). To accommodate more clusters, one can envision
running this approach recursively along with a divisive hierarchical
clustering scheme.

One might ask why we don't use a binary encoding of cluster membership
for $K>2.$ Indeed, this would be a thrifty use of qubits. Unfortunately,
in this case the objective function exceeds quadratic order in the
$s_{i}$, so it is not immediately amenable to solution on D-Wave
2X hardware. There exist methods for reducing the order \cite{reduction},
but at the expense of introducing additional qubits, the number of
which scales worse than the number in the one-hot encoding. The question
of whether the relaxed precision demands encountered using this order
reduction method gives some edge is left for future investigation.

\section{Results}

\subsection{One-hot encoding}

Here we present our clustering results obtained using one-hot encoding
as well as binary clustering technique. For problem sizes that fit
onto the hardware, results were obtained by running on actual QA hardware.
For larger instances which cannot fit, an open-source solver qbsolv
was used. Qbsolv was recently released by D-Wave \cite{Booth2009}
and allows QUBO problems that are too large to be embedded onto QA
devices to be solved. It is a hybrid approach that is designed to
be able to use the best of both classical and quantum optimization.
First, in the classical step, the large QUBO is divided into smaller
sub-QUBOs. Then each of these smaller sub-QUBOs can be solved
by QA. These solutions
are then stitched back together to find a full solution. In this paper qbsolv
is used as a purely classical technique using tabu search as a solver instead of QA.

\begin{figure}[htpb]
	%	\begin{center}
	\includegraphics[scale=0.50]{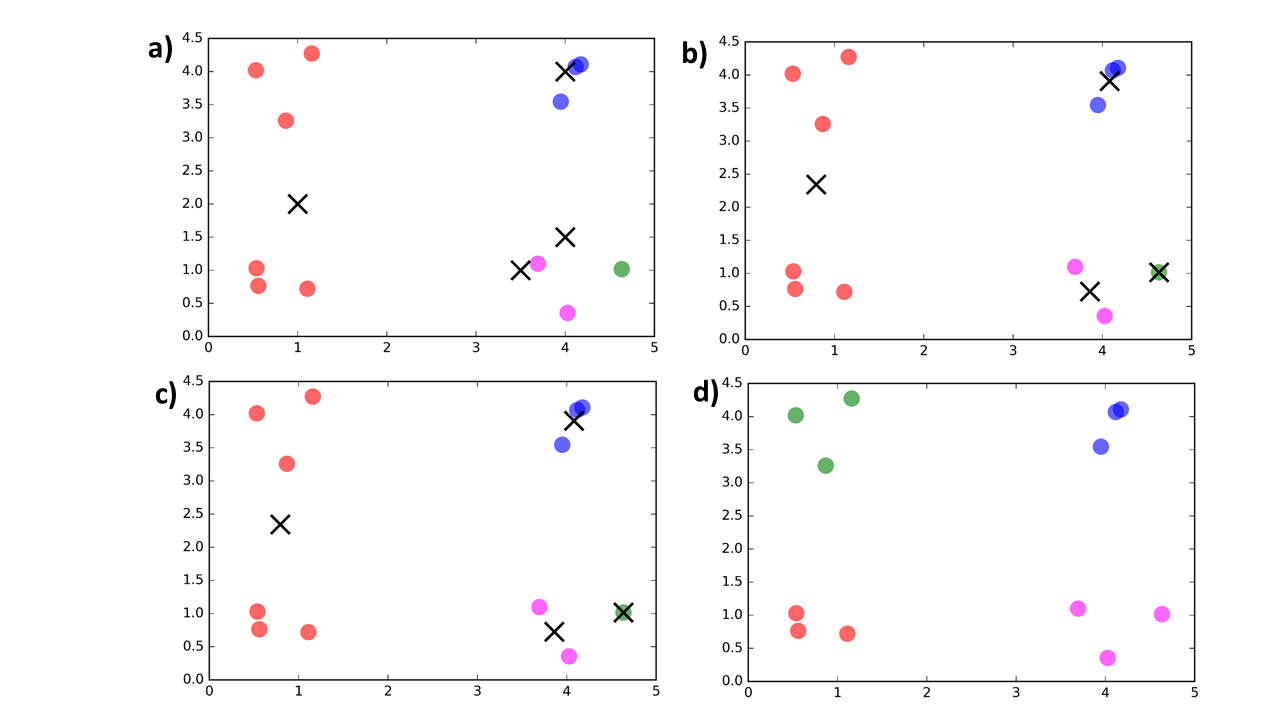}
	\caption{Clustering assignments obtained by $k$-means and one-hot
		encoding. The color coding represents clustering assignment. The black
		crosses indicate centroid position. a), b), c) indicate assignments
		obtained using $k$-means at iteration number 1, 2, and 20, respectively.
		d) indicates the assignment obtained using one-hot encoding run
		on the QA hardware.}
	\label{clust_fig1}
	%	\end{center}
\end{figure}

We use $k$-means clustering to compare performance. We first start
with a customized pedagogical problem to develop intuition of how
$k$-means can fail while the quantum-based approach succeeds. This
particular problem highlights the issue of improper initialization
of centroids during $k$-means which leads sub-optimal solutions.
Figure (\ref{clust_fig1}) a), b) and c) show the clustering of 12 points among four
clusters using $k$-means.

We show here a specific initialization of centroids. In this particular
configuration, where one centroid is centered in between a pair of
clusters, while the other three centroids are shared near the other
two true clusters. As can be seen, this can cause $k$-means to fail,
converging to the local minimum seen in Figure (\ref{clust_fig1} c). Note that such unfortunate
initializations occur more frequently when data points belong to $p>2$
dimensional space. Thus while this particular instance is trivially
solved by repeating with multiple random initializations, in higher
dimensions the problem becomes more common and severe. Figure 1d)
indicates clustering obtained using the one-hot encoding running on
QA hardware.

One-hot encoding is able to cluster points in a single step compared
to the iterative procedure followed by $k$-means. This instance required
48 variables encoded on the hardware which are partially connected
to each other. The appropriate embedding was found by the heuristic
embedding solver provided on the QA hardware. For all cases, QA hardware
was run with default parameters and post-processing was switched off.
Unless stated otherwise, in all cases 1000 samples were collected
and the spin configuration with lowest energy was selected as the
optimal solution. The couplings and biases were initially input in
QUBO form. For all instances, $d(x_{i},x_{j})$ was scaled to lie
within the range $[0,1]$. For all one-hot encoding instances, $\lambda$
was set equal to $N$ to satisfy the constraint \ref{eq:LambdaDef}.

\begin{figure}[htbp]
	\begin{center}
		\includegraphics[scale=0.50]{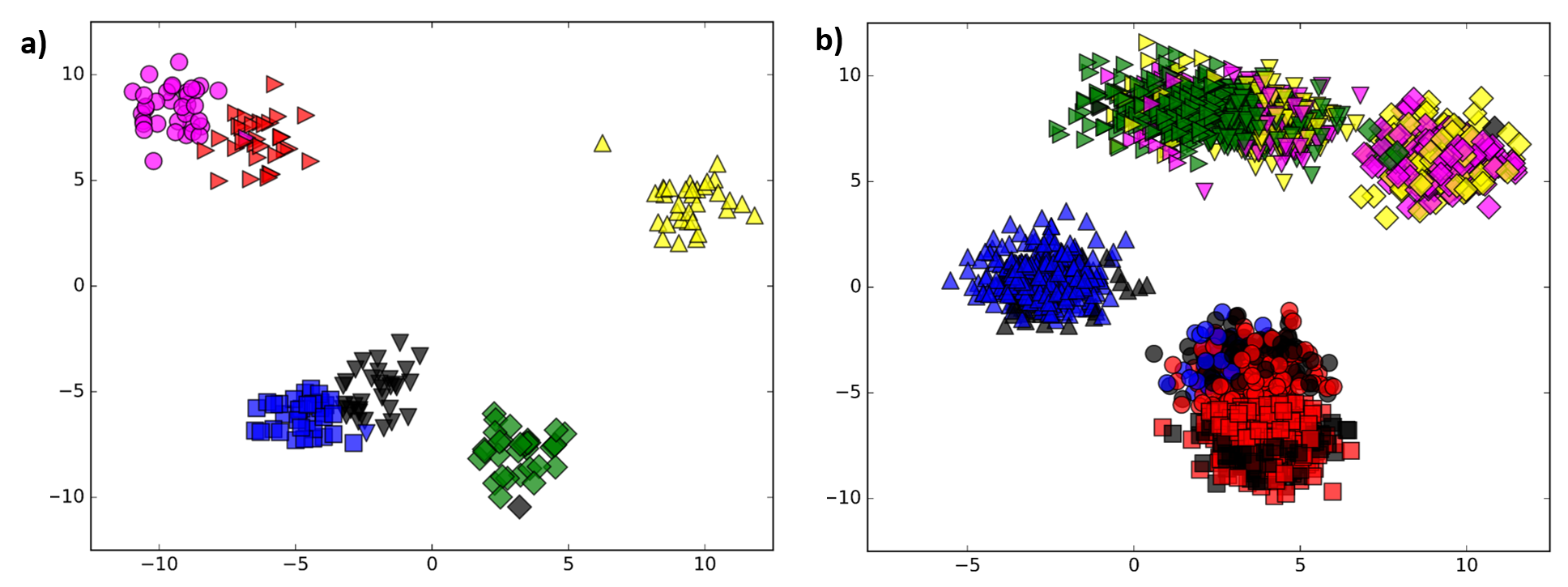}
		\caption{Clustering assignments obtained using $k$-means and one-hot
			encoding. The color coding represents cluster assignment obtained
			by one-hot encoding whereas marker shape represents cluster assignment
			obtained using $k$-means. a) and b) indicate $N=$ 200 and 2000
			case. One-hot encoding was run using qbsolv with \texttt{nrepeat}
			$=50$.}
		\label{clust_fig2}
	\end{center}
\end{figure}

We also used the one-hot encoding technique with qbsolv to solve larger
instances. We carried out clustering of $N=200,1000,$ and 2000 points
into $K=6$ clusters using one-hot and $k$-means. The points were
created as Gaussian blobs where overlap between clusters was allowed.
The scikit-learn \cite{scikit-learn} implementation of $k$-means
was used for comparison purposes. During $k$-means, the centroid
initialization was done randomly as well as using the $k$-means++
technique \cite{arthur2007k}. Ten initializations were used and the
one with lowest ``inertia'' was considered as the solution. Inertia
refers to the sum of distances between points and their respective
cluster centroid and is different from $W$. In $W$, each such distance
is weighted by the cluster size as well. $k$-means was considered
to have converged when the difference in inertia between successive
iterations reached $10^{-4}$ or 300 iterations were completed. Figure
(\ref{clust_fig2}) indicates the clustering assignments obtained using one-hot encoding
and $k$-means for $N=200$ and $2000$ case.

For $N=200$ case, one-hot encoding achieves assignments similar to
$k$-means. Whereas, the $N=2000$ case indicates major differences
between the assignments obtained using both the techniques. In order
to quantify these differences, in Table \ref{inertia_table} we compare the inertia
values obtained using one-hot encoding and $k$-means. 

\begin{table}
	\caption{Inertia values obtained using $k$-means and one-hot encoding}
	\label{inertia_table}
	%	\begin{center}
	\begin{tabular}{llll}
		\hline\noalign{\smallskip}
		N  & $k$-means++  & random  & one-hot encoding \\
		\noalign{\smallskip}\hline\noalign{\smallskip}  
		200  & 342.88  & 342.88  & 402.78 \\
		1000  & 1923.81  & 1923.84  & 4336.72 \\
		2000  & 3622.99  & 3622.99  & 15164.98 \\
		\noalign{\smallskip}\hline
	\end{tabular}
	%	\end{center}
\end{table}

The first two columns in Table \ref{inertia_table} refer to the two different ways
in which centroids were initialized during $k$-means. It is clear
that in all cases, $k$-means outperforms one-hot encoding. The performance
of $k$-means is much better than the one-hot encoding for $N=1000$
and $2000$ cases. This trend was somewhat expected as $N=1000$ and
$2000$ cases involve optimization of systems containing $6000$ and
$12000$ variables respectively. The quality of clustering obtained
using one-hot encoding depends heavily on the quality of solution
obtained using a given solver. We believe that the system sizes studied
here are simply too large for qbsolv to handle.

\begin{figure}[htbp]
	\begin{center}
		\includegraphics[scale=0.40]{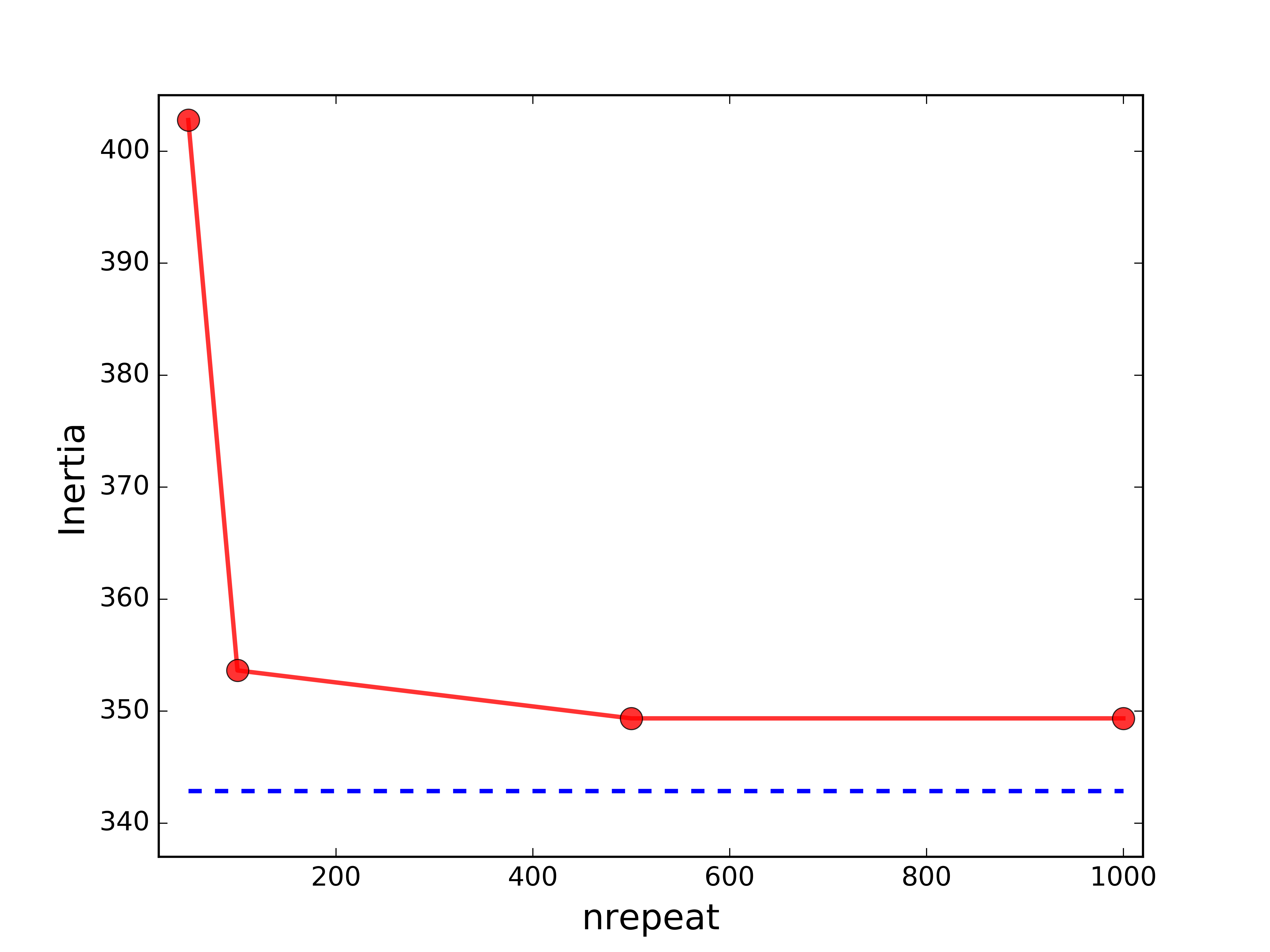}
		\caption{Evolution of inertia with \texttt{nrepeat} parameter. Red
			circles represent inertia values obtained using one-hot encoding and
			qbsolv. The red line serves as a guide to the eye. The dashed blue line indicates
			inertia value obtained using $k$-means.}
		\label{clust_fig3}
	\end{center}
\end{figure}

We used qbsolv as a blackbox in order to assess the performance of
proposed algorithms. In order to further examine the influence of
qbsolv's solution quality on clustering, we ran qbsolv with different
values of the parameter \texttt{nrepeat}. \texttt{nrepeat} is a hyperparameter
in the qbsolv solver. It is expected that setting \texttt{nrepeat}
to a higher value than the default value would return better quality
solutions. Figure (\ref{clust_fig3}) indicates the evolution of inertia with the parameter
\texttt{nrepeat} for $N=200$ case.

The clustering performance improves and gets closer to the $k$-means
performance as \texttt{nrepeat} value is increased. We obtained diminishing
improvements as \texttt{nrepeat} value was further increased. This clearly
indicates that as the solver gets better, the performance of the one-hot
encoding algorithm improves. We believe there are other qbsolv hyperparameters
which can be fine-tuned to achieve better results. A detailed study
is needed to understand the influence of hyperparameters on the quality
of solutions obtained using qbsolv.

\subsection{Binary Clustering}

\begin{figure}[htbp]
	\begin{center}
		\includegraphics[scale=0.40]{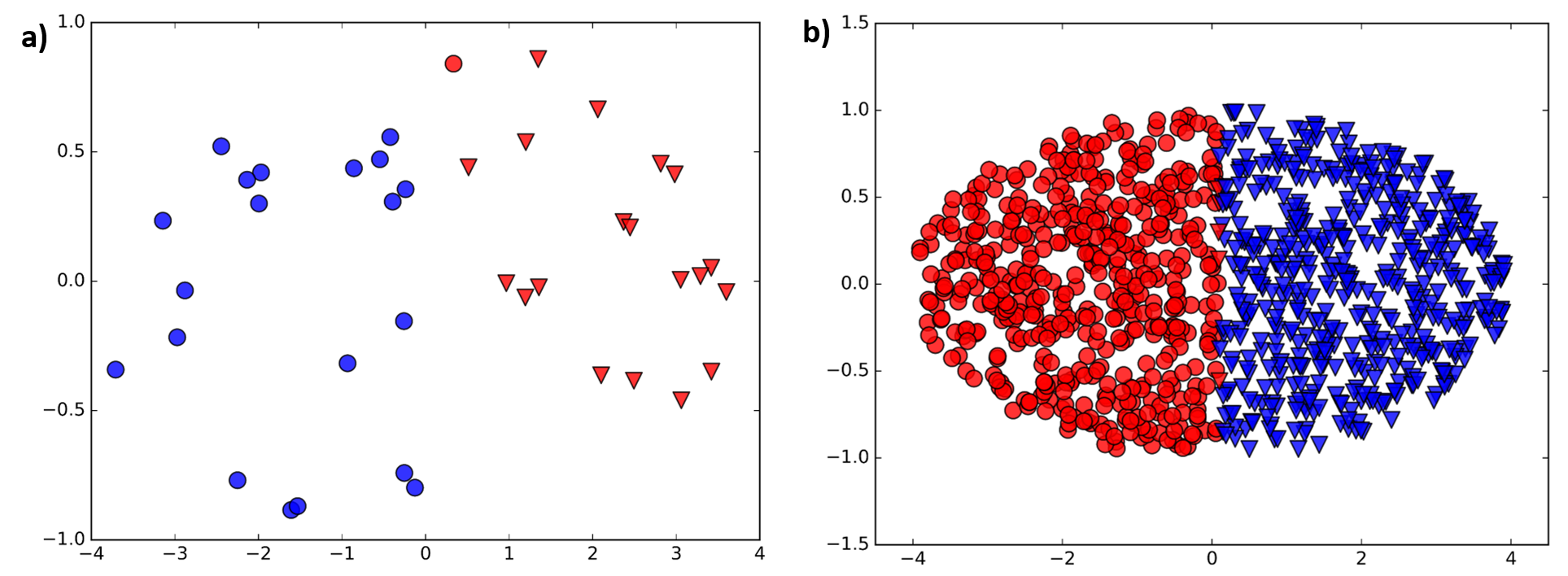}
		\caption{Clustering assignments obtained using $k$-means and binary
			clustering. The color coding represents the cluster assignment obtained
			by binary clustering whereas the marker shape represents the cluster assignment
			obtained using $k$-means. a) and b) show $N=200$ and 2000.}
		\label{clust_fig4}
	\end{center}
\end{figure}

We present our results for the binary clustering case. For binary
clustering we generate data points uniformly over an ellipse. Such
a distribution of points in two dimensions is known to be equivalent
to clustering of Gaussian distributed points in higher dimensions,
a fact that was used in \cite{savaresi2001performance} to compare
the performance of bisecting $k$-means and principal direction divisive partitioning (PDDP). Figure (\ref{clust_fig4}) indicates
a comparison between $k$-means and binary clustering method for $N=40$
and $2000$.

The binary clustering for $N=40$ was run on QA hardware. 40 fully
connected variables were embedded on the hardware using D-Wave's heuristic
embedding solver \cite{cai2014practical}. The $N=1000$ and $2000$
cases were treated using qbsolv. We observe few differences in the
cluster labels obtained using $k$-means and binary clustering. This
is evident from the inertia values obtained for both the techniques
tabulated in Table \ref{inertia_table2}.

\begin{table}
	\caption{Inertia values obtained using $k$-means and binary clustering.}
	\label{inertia_table2}
	%	\begin{center}
	\begin{tabular}{llll}
		\hline\noalign{\smallskip} 
		N  & $k$-means++  & random  & binary clustering \\
		\noalign{\smallskip}\hline\noalign{\smallskip} 
		40  & 52.81  & 52.81  & 52.89 \\ 
		1000  & 1330.90  & 1330.90  & 1330.94 \\ 
		2000  & 2784.07  & 2784.08  & 2784.14 \\
		\noalign{\smallskip}\hline 
	\end{tabular}
	% \end{center}
\end{table}

Inertia values obtained using binary clustering are very close to
those obtained using $k$-means. The performance of binary clustering
does not appear to get as low as the data size increases compared
to one-hot encoding.
\section{Discussion}

In this paper we introduced two formulations of the clustering objective
function which can be used to carry out clustering on quantum annealing
hardware. The one-hot encoding technique's performance was poor compared
to the $k$-means clustering on relatively large datasets. Our study
indicates that one-hot encoding yields better results as solver quality
improves. The need to use constraints severely limits the use of one-hot
encoding on QA hardware for clustering of all but very small datasets
due to the precision requirement. Moreover, one-hot encoding requires
substantially more qubits to encode the problem. The one-hot encoding
scheme adopted here, where each data point is associated to a $K$-bit
string, does not use the qubits efficiently and leads to cumbersome
constraint conditions. One could imagine a more efficient use of qubits
to avoid or decrease the dependency on constraints by using a binary
encoding-based scheme. In a binary encoding formulation, cluster assignment
for each data point can be represented by binary strings. Further
investigation in this direction is ongoing.

We tried to relax the constraint condition by setting $\lambda<(N-k)$.
As the constraint was relaxed we started observing violations where
a given point was assigned to more than one cluster. These solutions
were considered invalid in the current implementation. However there
are classes of fuzzy clustering algorithms where data points are allowed
to be assigned to more than one cluster. How such fuzzy frameworks
can be aligned with the current technique is not clear and requires
further investigation.

The binary clustering was observed to compare well with the $k$-means
results. The absence of constraints and use of fewer qubits makes
it particularly suitable for clustering of large datasets. These advantages
come with a limitation that binary clustering can only be used to
carry out binary splits at each step. We expect that in algorithms
such as divisive hierarchical clustering which makes binary split
at each step, one can benefit from the binary clustering technique.
Binary splitting in itself is an NP-hard problem and has traditionally
been treated with heuristic algorithms during divisive hierarchical
clustering \cite{Guenoche1991}. Hence, a hierarchical clustering
approach, in conjunction with binary clustering is expected to outperform
the current versions of divisive hierarchical clustering algorithms.
Any such comparison of the techniques introduced here with hierarchical
clustering and other advanced algorithms such as PDDP is left to a
future study. 

\begin{acknowledgements}
	We acknowledge the support of the Universities Space Research Association,
	Quantum AI Lab Research Opportunity Program, Cycle 2. 
	This is a pre-print of an article published in Quantum Information Processing. 
	The final authenticated version is available online
	at:https://doi.org/10.1007/s11128-017-1809-2
\end{acknowledgements}
\vspace{-12pt}

\bibliographystyle{unsrt}
\bibliography{BAB}

\providecommand{\noopsort}[1]{}\providecommand{\singleletter}[1]{#1}%
\begin{thebibliography}{10}

\bibitem{BenDor1999}
Amir Ben-Dor, Ron Shamir, and Zohar Yakhini.
\newblock Clustering gene expression patterns.
\newblock {\em Journal of computational biology}, 6(3-4):281--297, 1999.

\bibitem{Das2016}
Ranjita Das and Sriparna Saha.
\newblock Gene expression data classification using automatic differential
  evolution based algorithm.
\newblock In {\em Evolutionary Computation (CEC), 2016 IEEE Congress on}, pages
  3124--3130. IEEE, 2016.

\bibitem{gorzalczany2016}
Marian~B Gorza{\l}czany, Filip Rudz{\'\i}nski, and Jakub Piekoszewski.
\newblock Gene expression data clustering using tree-like soms with evolving
  splitting-merging structures.
\newblock In {\em Neural Networks (IJCNN), 2016 International Joint Conference
  on}, pages 3666--3673. IEEE, 2016.

\bibitem{marisa2013}
Laetitia Marisa, Aur{\'e}lien de~Reyni{\`e}s, Alex Duval, Janick Selves,
  Marie~Pierre Gaub, Laure Vescovo, Marie-Christine Etienne-Grimaldi, Renaud
  Schiappa, Dominique Guenot, Mira Ayadi, et~al.
\newblock Gene expression classification of colon cancer into molecular
  subtypes: characterization, validation, and prognostic value.
\newblock {\em PLoS Med}, 10(5):e1001453, 2013.

\bibitem{Xie2013}
Pengtao Xie and Eric~P. Xing.
\newblock Integrating document clustering and topic modeling.
\newblock {\em CoRR}, abs/1309.6874, 2013.

\bibitem{Balabantaray2015}
Rakesh~Chandra Balabantaray, Chandrali Sarma, and Monica Jha.
\newblock Document clustering using k-means and k-medoids.
\newblock {\em CoRR}, abs/1502.07938, 2015.

\bibitem{mudambi2002}
Susan Mudambi.
\newblock Branding importance in business-to-business markets: Three buyer
  clusters.
\newblock {\em Industrial marketing management}, 31(6):525--533, 2002.

\bibitem{sharma2013}
Arun Sharma and Douglas~M Lambert.
\newblock Segmentation of markets based on customer service.
\newblock {\em International Journal of Physical Distribution \& Logistics
  Management}, 2013.

\bibitem{chan2012}
Kit~Yan Chan, CK~Kwong, and Bao~Qing Hu.
\newblock Market segmentation and ideal point identification for new product
  design using fuzzy data compression and fuzzy clustering methods.
\newblock {\em Applied Soft Computing}, 12(4):1371--1378, 2012.

\bibitem{friedman2001elements}
Jerome Friedman, Trevor Hastie, and Robert Tibshirani.
\newblock {\em The elements of statistical learning}, volume~1.
\newblock Springer series in statistics New York, 2001.

\bibitem{hartigan1979}
John~A Hartigan and Manchek~A Wong.
\newblock Algorithm as 136: A k-means clustering algorithm.
\newblock {\em Journal of the Royal Statistical Society. Series C (Applied
  Statistics)}, 28(1):100--108, 1979.

\bibitem{johnson1967}
Stephen~C Johnson.
\newblock Hierarchical clustering schemes.
\newblock {\em Psychometrika}, 32(3):241--254, 1967.

\bibitem{jain2010}
Anil~K Jain.
\newblock Data clustering: 50 years beyond k-means.
\newblock {\em Pattern recognition letters}, 31(8):651--666, 2010.

\bibitem{Garey1979}
Michael~R. Garey and David~S. Johnson.
\newblock {\em Computers and Intractability: A Guide to the Theory of
  NP-Completeness}.
\newblock W. H. Freeman \& Co., New York, NY, USA, 1979.

\bibitem{papadimitriou1977}
Christos~H Papadimitriou.
\newblock The euclidean travelling salesman problem is np-complete.
\newblock {\em Theoretical computer science}, 4(3):237--244, 1977.

\bibitem{al1996computational}
Khaled~S Al-Sultana and M~Maroof Khan.
\newblock Computational experience on four algorithms for the hard clustering
  problem.
\newblock {\em Pattern recognition letters}, 17(3):295--308, 1996.

\bibitem{kirkpatrick1983}
Scott Kirkpatrick, C~Daniel Gelatt, Mario~P Vecchi, et~al.
\newblock Optimization by simulated annealing.
\newblock {\em science}, 220(4598):671--680, 1983.

\bibitem{selim1991}
Shokri~Z Selim and K1~Alsultan.
\newblock A simulated annealing algorithm for the clustering problem.
\newblock {\em Pattern recognition}, 24(10):1003--1008, 1991.

\bibitem{mitra1985}
Debasis Mitra, Fabio Romeo, and Alberto Sangiovanni-Vincentelli.
\newblock Convergence and finite-time behavior of simulated annealing.
\newblock In {\em Decision and Control, 1985 24th IEEE Conference on},
  volume~24, pages 761--767. IEEE, 1985.

\bibitem{szu1987}
Harold Szu and Ralph Hartley.
\newblock Fast simulated annealing.
\newblock {\em Physics letters A}, 122(3-4):157--162, 1987.

\bibitem{ingber1989}
Lester Ingber.
\newblock Very fast simulated re-annealing.
\newblock {\em Mathematical and computer modelling}, 12(8):967--973, 1989.

\bibitem{bouleimen2003}
KLEIN Bouleimen and HOUSNI Lecocq.
\newblock A new efficient simulated annealing algorithm for the
  resource-constrained project scheduling problem and its multiple mode
  version.
\newblock {\em European Journal of Operational Research}, 149(2):268--281,
  2003.

\bibitem{kadowaki1998}
Tadashi Kadowaki and Hidetoshi Nishimori.
\newblock Quantum annealing in the transverse ising model.
\newblock {\em Physical Review E}, 58(5):5355, 1998.

\bibitem{santoro2006}
Giuseppe~E Santoro and Erio Tosatti.
\newblock Optimization using quantum mechanics: quantum annealing through
  adiabatic evolution.
\newblock {\em Journal of Physics A: Mathematical and General}, 39(36):R393,
  2006.

\bibitem{denchev2016}
Vasil~S Denchev, Sergio Boixo, Sergei~V Isakov, Nan Ding, Ryan Babbush, Vadim
  Smelyanskiy, John Martinis, and Hartmut Neven.
\newblock What is the computational value of finite-range tunneling?
\newblock {\em Physical Review X}, 6(3):031015, 2016.

\bibitem{Born1928}
M.~{Born} and V.~{Fock}.
\newblock {Beweis des Adiabatensatzes}.
\newblock {\em Zeitschrift fur Physik}, 51:165--180, March 1928.

\bibitem{AlbashLidar2016}
T.~{Albash} and D.~A. {Lidar}.
\newblock {Adiabatic Quantum Computing}.
\newblock {\em ArXiv e-prints}, November 2016.

\bibitem{biamonte2016quantum}
J.~{Biamonte}, P.~{Wittek}, N.~{Pancotti}, P.~{Rebentrost}, N.~{Wiebe}, and
  S.~{Lloyd}.
\newblock {Quantum Machine Learning}.
\newblock {\em ArXiv e-prints}, November 2016.

\bibitem{dulny2016developing}
J.~{Dulny}, III and M.~{Kim}.
\newblock {Developing Quantum Annealer Driven Data Discovery}.
\newblock {\em ArXiv e-prints}, March 2016.

\bibitem{neven2009nips}
Harmut Neven, Vasil~S Denchev, Marshall Drew-Brook, Jiayong Zhang, William~G
  Macready, and Geordie Rose.
\newblock Nips 2009 demonstration: Binary classification using hardware
  implementation of quantum annealing.
\newblock {\em Quantum}, pages 1--17, 2009.

\bibitem{denchev2013binary}
Vasil~S Denchev.
\newblock {\em Binary classification with adiabatic quantum optimization}.
\newblock PhD thesis, Purdue University, 2013.

\bibitem{farinelli2016quantum}
Alessandro Farinelli.
\newblock A quantum annealing approach to biclustering.
\newblock In {\em Theory and Practice of Natural Computing: 5th International
  Conference, TPNC 2016, Sendai, Japan, December 12-13, 2016, Proceedings},
  volume 10071, page 175. Springer, 2016.

\bibitem{kurihara2009quantum}
Kenichi Kurihara, Shu Tanaka, and Seiji Miyashita.
\newblock Quantum annealing for clustering.
\newblock In {\em Proceedings of the Twenty-Fifth Conference on Uncertainty in
  Artificial Intelligence}, pages 321--328. AUAI Press, 2009.

\bibitem{sato2013quantum}
Issei Sato, Shu Tanaka, Kenichi Kurihara, Seiji Miyashita, and Hiroshi
  Nakagawa.
\newblock Quantum annealing for dirichlet process mixture models with
  applications to network clustering.
\newblock {\em Neurocomputing}, 121:523--531, 2013.

\bibitem{Ising1925}
Ernst Ising.
\newblock Beitrag zur theorie des ferromagnetismus.
\newblock {\em Zeitschrift f{\"u}r Physik}, 31(1):253--258, Feb 1925.

\bibitem{Dahl2013}
ED~Dahl.
\newblock Programming with d-wave: Map coloring problem (2013), 2013.

\bibitem{reduction}
H.~Ishikawa.
\newblock Transformation of general binary mrf minimization to the first-order
  case.
\newblock {\em IEEE Transactions on Pattern Analysis and Machine Intelligence},
  33(6):1234--1249, June 2011.

\bibitem{Booth2009}
Michael Booth, Steven~P. Reinhardt, and Aidan~year="2009" Roy.
\newblock Partitioning optimization problems for hybrid classical/quantum
  execution.

\bibitem{scikit-learn}
F.~Pedregosa, G.~Varoquaux, A.~Gramfort, V.~Michel, B.~Thirion, O.~Grisel,
  M.~Blondel, P.~Prettenhofer, R.~Weiss, V.~Dubourg, J.~Vanderplas, A.~Passos,
  D.~Cournapeau, M.~Brucher, M.~Perrot, and E.~Duchesnay.
\newblock Scikit-learn: Machine learning in {P}ython.
\newblock {\em Journal of Machine Learning Research}, 12:2825--2830, 2011.

\bibitem{arthur2007k}
David Arthur and Sergei Vassilvitskii.
\newblock k-means++: The advantages of careful seeding.
\newblock In {\em Proceedings of the eighteenth annual ACM-SIAM symposium on
  Discrete algorithms}, pages 1027--1035. Society for Industrial and Applied
  Mathematics, 2007.

\bibitem{savaresi2001performance}
Sergio~M Savaresi and Daniel~L Boley.
\newblock On the performance of bisecting k-means and pddp.
\newblock In {\em Proceedings of the 2001 SIAM International Conference on Data
  Mining}, pages 1--14. SIAM, 2001.

\bibitem{cai2014practical}
Jun Cai, William~G Macready, and Aidan Roy.
\newblock A practical heuristic for finding graph minors.
\newblock {\em arXiv preprint arXiv:1406.2741}, 2014.

\bibitem{Guenoche1991}
A.~Gu{\'e}noche, P.~Hansen, and B.~Jaumard.
\newblock Efficient algorithms for divisive hierarchical clustering with the
  diameter criterion.
\newblock {\em Journal of Classification}, 8(1):5--30, Jan 1991.

\end{thebibliography}

\end{document}